\documentclass[pre,a4paper,nofootinbib,twocolumn,showpacs,preprintnumbers,amsmath,amssymb,floatfix,amstex]{revtex4}
\newcommand{\beq}{\begin{equation}}         %
\newcommand{\eeq}{\end{equation}}           %
\newcommand{\beqa}{\begin{eqnarray}}            %
\newcommand{\eeqa}{\end{eqnarray}}          %
\newcommand{\ket}[1]{|\,#1\,\rangle}                %
\newcommand{\brac}[1]{\langle\,#1\,|}               %
\newcommand{\bra}[1]{\langle\,#1}               %
\newcommand{\vket}[1]{|\,#1\,)}                 %
\newcommand{\vbra}[1]{(\,#1}                     %
\newcommand{\vbrac}[1]{(\,#1\,|}             %
\newcommand{\grkbf}[1]{\mbox{\boldmath $#1$}}        %
\newcommand{\pd}{\partial}                           %
\newcommand{\dif}{\mbox{d}}                          %
\newcommand{\nx}[1]{\mathbf{#1}}                 %
\newcommand{\FP}{{\cal L}}               %
\newcommand{\hilN}{{\cal H}_{N}}             %
\newcommand{\hilNN}{{\cal H}_{N^2}}          %
\newcommand{\FPe}{{\cal L}_{\epsilon}}           %
\newcommand{\cDe}{{\cal D}_{\epsilon}}           
\newcommand{\De}{\text{\textbf{\textsf{D}}}_{\epsilon}}  
\newcommand{\Tqp}{\hat{T}_{(q,p)}}           %
\newcommand{\Tqpd}{\hat{T}_{(q,p)}^{^{\dagger}}}     %
\newcommand{\hrho}{\hat{\rho}}               %
\newcommand{\uhrho}{\hat{\rho}^{u}}               %
\newcommand{\shrho}{\hat{\rho}^{s}}               %
\newcommand{\hA}{\hat{A}}                %
\newcommand{\hB}{\hat{B}}                %
\newcommand{\hM}{\hat{M}}                %
\newcommand{\hR}{\hat{R}}
\newcommand{\hL}{\hat{L}}
\newcommand{\cqp}{c_{\epsilon}(q,p)}             %
\newcommand{\cmn}{\widetilde{c_{\epsilon}}(\mu,\nu)}     %
\newcommand{\qL}{\text{\textbf{\textsf{L}}}}         %
\newcommand{\SOp}[1]{\text{\textbf{\textsf{#1}}}}    
\newcommand{\doll}{\text{\textbf{\textsf{L}}}_{\epsilon}}
\newcommand{\II}{\hat{I}}        
\newcommand{\TT}{\mathbb{T}^{2}}        
\newcommand{\RR}{\mathbb{R}^2}          
\newcommand{\CC}{\mathbb{C}}
\newcommand{\Ldos}{\mathbb{L}^2}        
\newcommand{\hU}{\hat{U}}
\newcommand{\hV}{\hat{V}}
\newcommand{\hUd}{\hat{U}^\dagger}              %
\newcommand{\uxu}{\hU\otimes \hU^{\dagger}}         %
\newcommand{\ie}{\mbox{{\em i.e.\/}} }          %
\newcommand{\etal}{\mbox{{\em et al.\/}} }      %
\newcommand{\tr}{\mbox{Tr}}             %
\newcommand{\NtoOO}{N\to\infty}             %
\newcommand{\etoO}{\epsilon\to 0}           %
\newcommand{\equa}[1]{Eq.~(\ref{#1})}           %
\newcommand{\fig}[1]{Fig.~\ref{#1}}         %

\newcommand{\rmd}{{\rm d}}
\newcommand{\rmi}{{\rm i}}
\newcommand{\rme}{{\rm e}}
\newcommand{\nmax}{n_{\text{max}}}
\newtheorem{prop}{Proposition}
\usepackage{graphicx,graphics,wrapfig,rotating}     
\usepackage{dcolumn}                    
\usepackage{bm,fancybox}                
\usepackage{times,euscript,eufrak,oldgerm}                  %
\begin{document}
\title{ Spectral properties and classical
decays in quantum open systems} %
\author{Ignacio Garc\'\i a-Mata}
\email{garciama@tandar.cnea.gov.ar}
\author{Marcos Saraceno}
\email{saraceno@tandar.cnea.gov.ar}
\affiliation{%
Dto. de F\'\i sica, Comisi\'on Nacional de Energ\'\i a At\'omica.
Av. del Libertador 8250 (1429), Buenos Aires, Argentina.
}%
\date{\today}%
\begin{abstract}
We study the relationship between the spectral properties of
diffusive open quantum maps and the classical spectrum of
Ruelle-Pollicott resonances. The leading resonances determine
 the asymptotic time regime for several quantities of interest
- the linear entropy, the
Loschmidt echo and the correlations of the initial state. A
numerical method that allow an efficient calculation of the
leading spectrum is developed using a truncated basis adapted to
the dynamics.
\end{abstract}
\maketitle

\section{Introduction}
 The study of the emergence of classical features in systems ruled
 by quantum mechanics is as old as quantum mechanics itself. When
 the quantum system is isolated and the evolution unitary, these
 features appear in the WKB semiclassical limit, which is of
 paramount importance in establishing the quantum classical
 correspondence in integrable systems. In its more modern form,
 the EBK quantization rule \cite{EBK}, it shows the direct
 connection between tori in phase space and quantized
 eigenfunctions in Hilbert space. In chaotic systems, the
 relationship is more subtle and is embodied in the celebrated
 Gutzwiller trace formula \cite{Gutzwiller}, relating sets of unstable
 periodic orbits to the density of states. The limits of applicability of
 these semiclassical methods and the insight they provide on the quantum dynamics
 of isolated chaotic systems has inspired most of the recent
 research in the area of quantum chaos.

 In open quantum systems, on the contrary, the emergence of
 classical features has been studied mainly in the time evolution
 of a different set of observables, most notably the rate of linear
 entropy growth (or purity decay)\cite{zurek}, the Loschmidt
 echo or fidelity \cite{peres,losch,prosen}
 and the decay of correlations \cite{Srednicki}. These studies have demonstrated
 that in certain well defined regimes for chaotic systems the classical
 Lyapunov exponent governs these rates and that the evolution of
 localized quantum densities in phase space becomes classical.

In this article we consider this question from the point of view
of the spectral properties of the classical and quantum
propagators. Classical densities evolve according to the Liouville
equation whose solution can be written in terms of a propagator called 
Perron-Frobenius operator(PF)\cite{PF}. It is unitary on $\Ldos$. However, for chaotic
systems, correlation functions exhibit oscillations and exponential decay. The
decay rates are given by the poles of the resolvent of the PF
operator, the so-called Ruelle-Pollicott (RP) resonances\cite{ruel}. By limiting the
resolution of the functional space, one can effectively truncate the PF to a
nonunitary operator of finite size (say $N\times N$) with a spectrum lying
entirely inside the unit circle,
except for the simply degenerate eigenvalue 1. In the, properly taken, limit
of infinite size and no coarse graining, the isolated eigenvalues 
turn out to be the RP resonances\cite{blank,nonn}. 
As shown in \cite{garma}, the linear entropy and the Loschmidt echo, 
for asymptotic times much longer than the Ehrenfest time\footnote{In our case
the Ehrenfest time $n_E$ is related to the time it takes for an initially localazed
package to reach the borders of phase space due to exponential instability (it
is sometimes called ``log time'').}, also
show characteristic decay rates governed by the classical
RP resonances\cite{ruel}.
Experimental evidence of this dependence on RP 
resonances was observed for the
first time in \cite{sridhar}.
Our approach is similar in spirit to the calculations performed on the sphere
for the dissipative kicked top by Haake and collaborators
\cite{haak,webe,braun}, Fishman \cite{susy,khod} and, for the baker's
map, by Hasegawa and Saphir \cite{hase}.
We model the unitary dynamics by means of a quantum map and implement a diffusive
superoperator represented by a Kraus sum. Two recent works by Blank,
\etal\cite{blank} and
Nonnenmacher\cite{nonn} provide a rigorous
theoretical underpinning to our calculations for quantum and classical
maps on the torus.

The plan of the paper is as follows. Section II provides a short
account of the
 quantization procedure for maps acting on a classical surface with
 periodic boundary conditions in both coordinates and momenta, \ie a torus. 
In Sec.III
we implement the open system dynamics with the definition 
of a diffusion superoperator represented as a Kraus sum. 
The general spectral properties of both the unitary and the noisy part, 
as well as those of the combined action are studied. 
Sec.IV deals with the relationship between the classical and the 
quantum resonances and, utilizing recently proved theorems 
\cite{blank, nonn}, how they coincide in specific ranges of 
$\hbar$ and of the noise strength. As a consequence we show
that the asymptotic time behaviour of several quantities 
is classical and depends on the Ruelle-Pollicott resonaces closer to the unit circle. 
A numerical method that
allows the calculation of the leading spectrum of resonances is developed.
Sec.V illustrates this correspondence taking the perturbed Arnold cat map as an example.
We relegate to the Appendix some notation concerning the spectral decomposition of
superoperators and the details of the numerical method.
\section{Unitary dynamics on the torus $\TT$}
        \label{sec:phspace}
 We picture the classical phase space as a square of unit area with sides identified.
 The classical transformations will map this square onto itself, thus providing
 a simple model of Hamiltonian area preserving dynamics. The fact that the phase space
 has finite area brings some well known special features to the quantization that we
 briefly review. Refs. \cite{ozorivas,Miquel} provide a more extensive account.

\subsection{ The Hilbert space}
    \label{sec:Hspace}
As the phase space  has finite area, which we normalize to unity, the Hilbert
space  $\hilN$ is finite and its dimension $N$ sets the value of Planck's constant
 to $\hbar=(2\pi N)^{-1}$. The position and momentum bases are then sets of discrete states
$\ket{q},\ket{p}, ~ q,p=0,..N-1$ which are related by the discrete Fourier
transform (DFT) of dimension $N$
\begin{equation}
    \label{eq:ft}
\bra{p}\ket{q}={1\over\sqrt{N}}\,\text{e}^{-{2\pi \rmi\over N}pq}.
\end{equation}

A vector $\ket{\varphi(t)}$ in $\hilN$ characterizes pure states 
of the system and can be represented by the amplitudes 
$\langle q|\varphi\rangle ,\langle p| \varphi\rangle $ 
in the coordinate or momentum basis, respectively.

In the description of open systems it is imperative to represent
states by a density operator $\hrho$. They  form a subset of
self-adjoint, positive semidefinite matrices with unit trace in
\mbox{$\hilNN\stackrel{\mbox{\footnotesize
def}}{=}\hilN\otimes\hilN^{*}$} , the space of complex $N\times N$
matrices, usually called in this context Liouville
 space. While Hilbert space is the natural arena for unitary dynamics,
 this much larger Liouville space
  sets the stage for the more general description of open quantum dynamics.
It acquires the structure of a Hilbert space with the usual
introduction of the matrix scalar product
\begin{equation}
	\label{eq:scprod}
(\hA,\hB)=\tr(\hA^\dagger\hB).
\end{equation}
where $\hA,\hB\in \hilNN $. Linear transformations in this space
are termed superoperators, they
map operators into operators and are represented by $ N^2 \times
N^2 $ matrices. In Appendix A we review the various notations and
properties related to this space.
\subsection{Translations on the torus}
    \label{sec:translations}
  The usual translation operator in the infinite plane $\RR$ is
\begin{eqnarray}
    \label{eq:Tqp}
\hat{T}_{(q,p)}&=& \text{e}^{-{i\over \hbar}(q\hat{P}-p\hat{Q})}\\
         &=& \text{e}^{-{i\over \hbar}q\hat{P}}
             \text{e}^{{i\over \hbar}p\hat{Q}}
         \text{e}^{{i\over 2\hbar}q p}\nonumber \\
         &=& \hat{U}^{q}\hat{V}^{p}\text{e}^{{i\over 2\hbar}q p},
\end{eqnarray}
where $\hU$ and $\hV$ generate shifts in the position and momentum eigenbasis
respectively.
 On the torus the main difference  is that
 the infinitesimal translation operators $\hat P,\hat Q$ with the usual
 commutation rules cannot be defined because
 position and momentum eigenstates are discrete.
 However, {\em finite\/} translation operators $\hat{U}$
and $\hat{V}$ that have the property
\begin{equation}
    \label{eq:UV}
\hat{V}^{p}\hat{U}^{q}=\hat{U}^{q}\hat{V}^{p}\text{e}^{i{2\pi\over
N}qp}
\end{equation}
can be defined and they generate finite cyclic shifts in the
respective bases \cite{schwinger}.
 The $N\times N$ grid of coordinate and momentum states constitutes
the quantum phase space for the torus.
Eq.~(\ref{eq:UV}) allows a
definition of a translation operator $\hat{T}_{(q,p)}:\hilN
\to\hilN$, $q,p$ integers, analog to \equa{eq:Tqp}. The
action of $\hat{T}$ on position and momentum eigenstates is
\begin{eqnarray}
\hat{T}_{(q_1,p_1)}\ket{q}&=&\exp\left[i {2\pi\over N}p\left(q+{q_1\over
2}\right)\right]\ket{q+q_1}\\
\hat{T}_{(q_1,p_1)}\ket{p}&=&\exp\left[-i {2\pi\over
N}q\left(p+{p_1\over 2}\right)\right]\ket{p+p_1}.
\end{eqnarray}
These equations confirm that $\hat{T}_{(q_1,p_1)}$ are indeed
phase space translations. They satisfy the Weyl group composition
rule
\begin{equation}
\label{eq:weyl}
\hat{T}_{(q_1,p_1)}\hat{T}_{(q_2,p_2)}=\hat{T}_{(q_1+q_2,p_1+p_2)}
\text{e}^{i{\pi\over N}(p_1 q_2-q_1 p_2)}.
\end{equation}
The $N^2$ translations $\hat{T}_{(q,p)} , p,q=0,..N-1$ satisfy the
orthogonality relation
\begin{equation}
\tr(\hat{T}^{\dagger}_{(q,p)}\hat{T}_{(q',p')})= N\delta_{qp,q'p'}
\end{equation}
thus constituting an orthogonal basis for the Liouville space
$\hilNN$.

The expansion of any operator $\hA$ in this basis constitutes the
chord \cite{ozorivas} or characteristic function representation.
This representation assigns to every $\hA \in \hilNN$ the c-number
function $a(q,p)= N^{-1}\,\tr ( \hA \Tqpd)$ and therefore every
operator has the expansion
\begin{equation}
\label{eq:chord}
 \hA=\sum_{q,p}a(q,p)\Tqp.
\end{equation}

For representation purposes we also use a basis of ``phase point '' operators
that constitute the Weyl, or center \cite{ozorivas}, representation. In this
basis the density operator is the discrete Wigner function of the quantum state.
The peculiar features of the discrete Wigner function for Hilbert spaces of
finite dimension have been described recently in \cite{Miquel}.
\subsection{Unitary Dynamics: quantum maps}
        \label{sec:unitary}
A classical map is a dynamical system that usually, but not
exclusively, arises form the discretization of a continuous time
system (by means of a Poincar\'e section, for example). Although
it is always possible, by integration of the equations of motion,
to derive the map from a Hamiltonian, this connection is rather
involved and in many instances it is more useful to  model
specific features of Hamiltonian dynamics by directly specifying
the map equations without going through the integration step. The
same is true in quantum mechanics: instead of modeling the
Hamiltonian operator and integrating it to obtain the unitary
propagator, it is simpler to model directly the unitary map.
Classically an area preserving map is characterized by a finite
canonical transformation and the corresponding quantum map is the
unitary propagator that represents this canonical transformation.
There are no exact and systematic procedures to realize this
correspondence. On the 2-dimensional plane $\RR$ relatively
standard procedures (see \cite{OdAbook}) give an approximation of
the propagator in the semiclassical limit as
\begin{equation}
U(q_1,q_2)=\left({i\over\hbar}{\pd^2 S\over\pd q_1\pd q_2}\right)^{1/2}
\exp\left[{i\over \hbar}S(q_2,q_1)\right],
\end{equation}
where $S$ is the action along the unique classical path from $q_1$
to $q_2$, and where  for simplicity  we do not consider the
existence of multiple branches and Maslov indices. Only for linear
symplectic maps on $\RR$ this unitary propagator is exact, and
then
 $S$ is minus
the quadratic generating function of the linear transformation.
On the other hand, several ad-hoc procedures for the quantization of
specific maps have been devised:
some integrable (translations\cite{schwinger} and
shears) and chaotic maps, such as cat maps \cite{hannay,OdA},
baker maps \cite{balazs,saraceno} and the standard map \cite{izra}
. Also all maps of a ``kicked'' nature, realized as compositions
of non-commuting nonlinear shears can be quantized, as well as
 periodic time dependent  Hamiltonians \cite{BerryVoros}.
Once the quantum propagator has been constructed, the advantages
of using quantum maps to model specific features of quantum
dynamics become apparent. The propagator $\hU$ is a unitary
$N\times N$ matrix, propagation of a pure state is achieved simply
by matrix multiplication, and finally  the classical limit is
obtained by letting $N\to\infty$.

In Liouville space the evolution of the density operator $\hrho$
by the map $\hU$ is given by
\begin{equation}
    \label{eq:rho}
\hrho'=\hU\rho\hU^{\dagger}.
\end{equation}
As a linear map acting on $\hilNN$  \equa{eq:rho} can be written as
\begin{equation}
    \label{eq:uevol}
\hrho'=\uxu(\hrho)=\SOp{U}(\hrho),
\end{equation}
  In what follows the notation $\uxu $ is meant to be equivalent to
 the $Ad(U)$ notation customary in group theory.
  The linear operator
$\SOp{U}\stackrel{\mbox{{\footnotesize def}}}{=}\uxu$ is a unitary
$N^2\times N^2$ matrix.
\section{Noisy Dynamics}
    \label{sec:noisy}
Realistic quantum processes always involve a certain degree of
interaction between system and environment. In this case the
evolution of the system
 is not unitary and requires a description in Liouville space. This loss
 of unitarity  leads to decoherence and to the emergence of classical
  features \cite{zurekRMP03,pazLH} in the evolution.
When the environment is taken into account the evolution of the
system is governed by a {\em master equation\/}, which takes the
form of a hierarchy of integro-differential equations. A drastic
simplification follows from the assumption that the environment
reacts to the system sufficiently fast, in such a manner that the
system looses all prior memory of its state, \ie that the
evolution is Markovian. \footnote{For a detailed description of
quantum noise and quantum Markov processes see
\cite{gardiner,buchleitner}.}. The resulting Lindblad equation
\cite{L,Perc}
\begin{eqnarray}
\frac{\rmd\hat\rho}{\rmd t}&=&-{\rmi\over\hbar}[\hat
H,\hat\rho]+\nonumber \\                                 %
& &\mbox{\hspace{0.25cm}} +{1\over\hbar}\sum_j(\hat
L_j\hat\rho\,\hat L_j^\dag -\case12\hat L_j^\dag\hat L_j\hat \rho
-\case12\hat\rho\,\hat L_j^\dag\hat L_j). \label{Lind}
\end{eqnarray}
determines the evolution of open quantum systems through a
Hamiltonian $\hat H$ that governs the unitary noiseless evolution
and the Lindblad $\hat L_i$ operators that model the interaction
with the environment. The particular structure of the equation
ensures that the evolution preserves the total probability, the
positive semi-definiteness and hermiticity of the density matrix.
The infinitesimal propagator is a linear operator in Liouville
space which can be integrated to yield a finite linear mapping
\begin{equation}
    \label{eq:sevol}
\hrho=\SOp{S}(\hrho_0).
\end{equation}
This mapping, as a reflection of the structure of the Lindblad
equation, also has a particular form that guarantees the
preservation of the general properties of the density operator.
The general form, called the Kraus representation \cite{Kraus} is
\begin{equation}
    \label{eq:kraus}
\SOp{S}(\hrho)=\sum_{\mu}\hM_{\mu}\hrho \hM^{\dagger}_{\mu}
=\Bigg[\sum_{\mu}\hM_{\mu}\otimes
\hM^{\dagger}_{\mu}\Bigg](\hrho).
\end{equation}
The only further restriction on the $\hM_\mu$ operators arises
from the preservation of the trace that requires
\begin{equation}
    \label{eq:tr-pres}
\sum_{\mu}\hM_{\mu}^{\dagger}\hM_{\mu}=\II.
\end{equation}
In what follows we will select them from a certain
complete family, with a specific norm. In that case the
representation takes a more general form
\begin{equation}
    \label{eq:kraus1}
\SOp{S}(\hrho)=\sum_{\mu}c_\mu \hM_{\mu}\hrho \hM^{\dagger}_{\mu}
\end{equation}
where now the positivity requirements are $c_\mu\geq 0$ and
\begin{equation}
    \label{eq:tr-pres1}
\sum_{\mu}c_\mu \hM_{\mu}^{\dagger}\hM_{\mu}=\II.
\end{equation}
Within this general framework, just as in the case of quantum
maps, we have the choice of modeling the noise through the
Lindblad operators or directly in terms of the integrated form via
the Kraus operators. In what follows we choose the latter and thus
we model the evolution by specifying a quantum map to represent
the unitary evolution followed by a noisy step, modeled by  its
Kraus superoperator form. An evolution of the density matrix
specified in this way is known in the literature
\cite{preskill,chuang,schumacher,caves} as a {\em quantum
operation}. It includes the special case of unitary evolution when
the sum is limited to only one term. In that case, and only then,
the dynamics is reversible. A general superoperator has no
inverse.
\subsection{Quantum coarse grained dynamics}
    \label{sbsec:QCGD}
We are interested in modeling the effect of a small amount of
noise on the evolution of an otherwise unitary quantum map
\cite{bian,nonn,braun,garma}. We assume that the one step
propagator results from the {\em composition\/} of two
superoperators. The first is the unitary propagator $\SOp{U}$ and
the second is a quantum diffusion superoperator $\De$, defined by
\begin{equation}
    \label{eq:deps}
\De=\sum_{q,p}\cqp \Tqp\otimes\Tqpd,
\end{equation}
that introduces decoherence.
The linear form of the full propagator is
\begin{equation}
    \label{eq:superop}
\doll=\De\circ\SOp{U}
\end{equation}
 and its action on a density matrix $\hrho$ is 
\begin{equation}
    \label{eq:superop1}
\doll(\hrho) =\sum_{q,p}\cqp \Tqp\hU\hrho\hUd\Tqpd.
\end{equation}
As the Kraus operators in this case are unitary the condition
(\ref{eq:tr-pres1}) becomes simply
\begin{equation}
\sum_{q,p}\cqp=1.
\end{equation}
Subject to this condition $\cqp$ can be an arbitrary positive function of
$q$ and $p$. Its significance in terms of coarse graining is
clear:  as long as $\cqp$ is
peaked around $(0,0)$ and of width $\epsilon$, the action that follows
the unitary step consists in
displacing the state incoherently over a phase space region of
order $\epsilon$. To avoid a net drift in any particular
direction, $\cqp$ must be an even function of the arguments $q$
and $p$. From this imposition and the fact that
\begin{equation}
\Tqpd=\hat{T}_{(-q,-p)}
\end{equation}
it follows, from the properties of the matrix scalar product 
(\ref{eq:scprod}) that $\De$ is hermitian
(see App.~\ref{ap:vec} for details on the scalar product).

The spectral properties of the separate superoperators $\SOp{U}$
and $\De$ are simple to obtain. If the Floquet spectrum of the quantum map
is
 \begin{equation}
 \hU\ket{\phi_k}=\rme^{\rmi\xi_k}\ket{\phi_k}
\end{equation}
 then the spectrum of $\SOp{U}$ is unitary and given by
\begin{equation}
\SOp{U}(\ket{\phi_k}\bra{\phi_j}|)=
      \rme^{\rmi(\xi_k-\xi_j)}\ket{\phi_k}\bra{\phi_j|}.
\end{equation}
 To obtain the spectrum of $\De$ we use the composition
 rule (\ref{eq:weyl}) to show that
\begin{equation}
\hat{T}_{(q,p)}\hat{T}_{(\mu,\nu)}\hat{T}_{(q,p)}^\dagger
    =\exp\left[\rmi\frac{2\pi}{N} (\nu q-\mu p)\right]\hat{T}_{(\mu,\nu)}.
\end{equation}
We then derive
\begin{eqnarray}
\De\left(\hat{T}_{(\mu,\nu)}\right)
        &=&\sum_{q,p}\cqp\hat{T}_{(q,p)}\hat{T}_{(\mu,\nu)}
\hat{T}_{(q,p)}^{^\dagger}\nonumber \\
        &=&\sum_{q,p}\cqp \exp\left[\rmi {2\pi\over N}
(\nu q-\mu p)\right]\hat{T}_{(\mu,\nu)}\nonumber \\
        &=&\cmn\hat{T}_{(\mu,\nu)}.
\end{eqnarray}
Therefore, the $N^2$ eigenvalues of $\De$ are given by the $2D$
discrete Fourier transform  $\cmn$ of the coefficients $\cqp$.
The eigenfunctions are the translation
operators themselves.
Hence, using the bra-ket notation described in
appendix~\ref{ap:vec}, the spectral decomposition of $\De$ is
\begin{equation}
    \label{eq:deps2}
\De=\sum_{\mu,\nu}\Big|T_{(\mu,\nu)}\Big)\cmn\Big(T_{(\mu,\nu)}\Big|,
\end{equation}
in analogy with \equa{eq:spectrald}.

Physically, the action of $\De$ is quite simple in the chord
representation (\ref{eq:chord}): if $\hrho$ is expanded as

\begin{equation}
 \hrho=\sum_{\mu,\nu}\rho_{\mu,\nu}\hat{T}_{(\mu,\nu)}
\end{equation}
 then
\begin{equation}
\De(\hrho)=\sum_{\mu,\nu}\cmn\rho_{\mu,\nu}\hat{T}_{(\mu,\nu)}.
\end{equation} Thus the coefficients in the chord representation are
suppressed selectively according to $\cmn$.

%
\begin{figure}[!htb]
\scalebox{0.75}{
\includegraphics*[180,346][429,451]{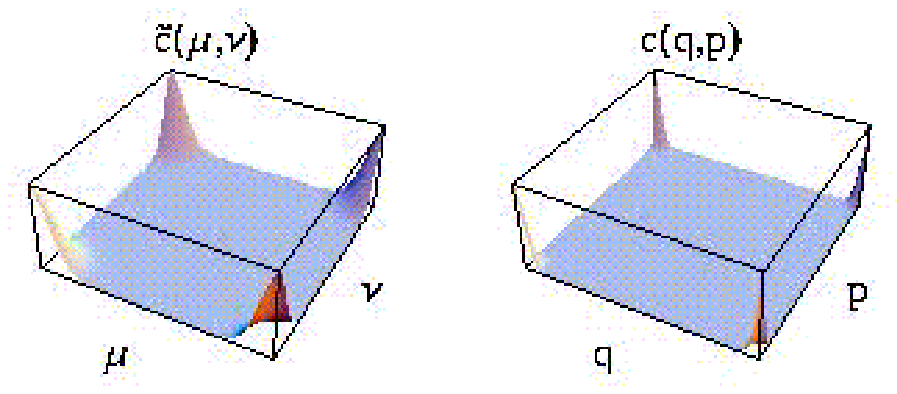}}
\caption{The left pane $\cmn$ shows the eigenvalues of $\De$ for
$\epsilon=0.15$ and $N=100$. The DFT of this function generates
the coefficients $\cqp$ (right pane) of the Kraus representation
of $\De$ of \equa{eq:deps}. \label{fig:fourier}}
\end{figure}
%

It is evident from Eqs.~(\ref{eq:deps}) and (\ref{eq:deps2}) that
the diffusion superoperator thus defined can be specified
indistinctly either by $\cqp$ or by $\cmn$ . For an efficient
numerical implementation of its action we have found
convenient to specify the latter as
\begin{equation}
\cmn=\text{e}^{-{1\over 2}\left({\epsilon N\over\pi}\right)^2\left(
\sin^2[\pi\mu/N]+\sin^2[\pi\nu/N]\right)},
\end{equation}
This is a smooth Gaussian-like periodic function of the integer
variables $\mu$ and $\nu$. For large values of $N$ it is very
close to the Gaussian
\begin{equation}
\widetilde{c}(\mu,\nu)\simeq \text{e}^{-\epsilon^2 (\mu^2 +\nu^2)/2}
\end{equation}
This means that the action of $\De$ will leave essentially
unaltered the coefficients $\rho_{\mu,\nu}$ in a region of size
 $\sim 1/(\epsilon N)$ (\fig{fig:fourier}, left) around the origin
 while strongly suppressing those outside.  The
backward DFT of $\cmn$ does not have a simple analytic expression
but from general properties of the DFT it will also be a Gaussian
like function with the complementary width $\sim\epsilon/2\pi$
 (\fig{fig:fourier},right).

\begin{figure}[!htb]
\begin{center}
\scalebox{0.55}{
\includegraphics*[80,548][524,776]{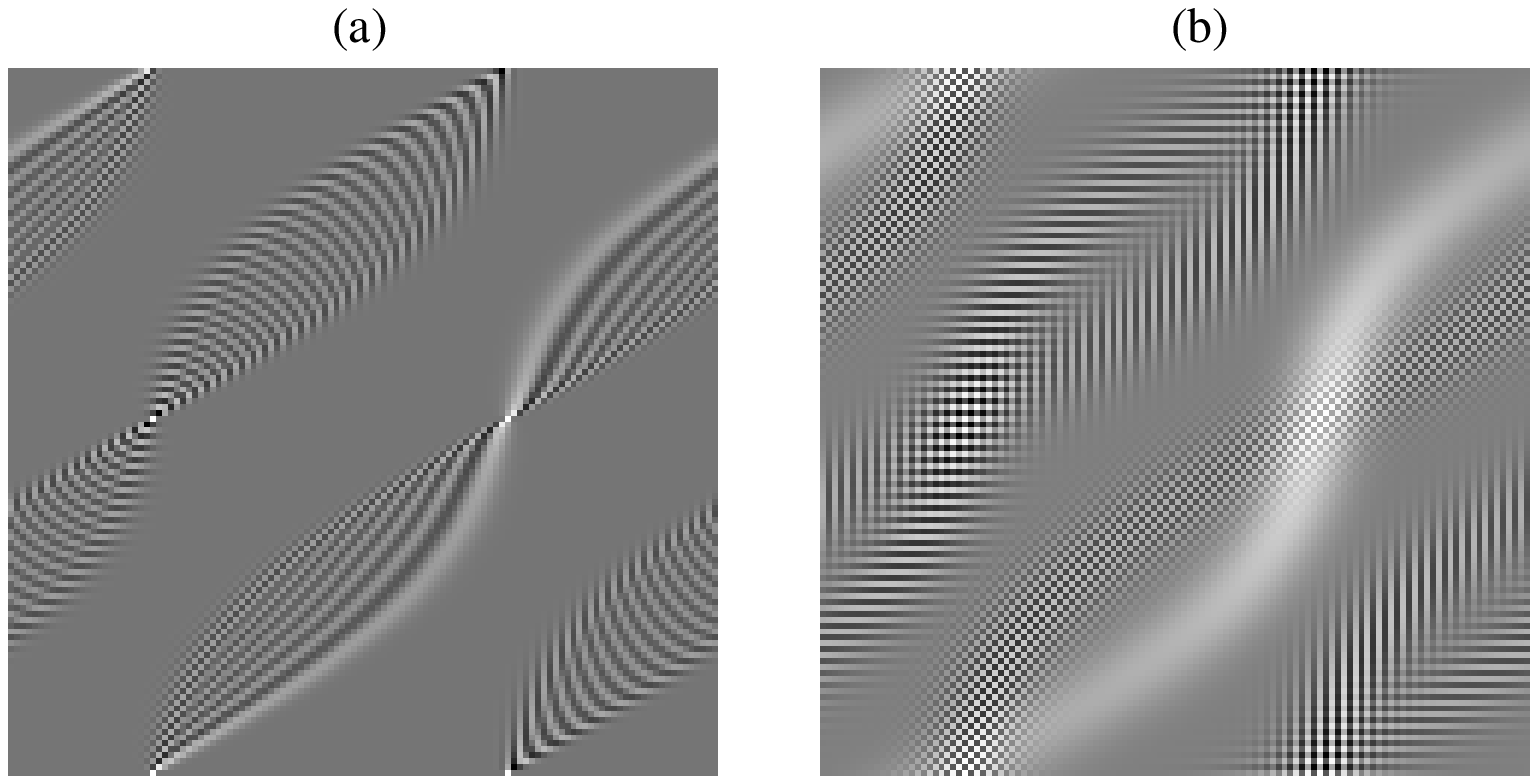}}
\caption{Display of the action of $\De$. Panel (a) shows the
Wigner function after the step $\SOp{U}$ has been applied to a
position state ($\rho_0=\ket{q_0}\brac{q_0}$). Panel (b) shows the
state after the full propagator $\doll=\De\circ\SOp{U}$ has acted.
The map is the perturbed cat map of \equa{eq:pcat}$k=0.02$,
$N=60$, $\epsilon=0.25$.
\label{fig:onestepw}}
\end{center}
\end{figure}
The action of $\De$ progressively washes out the quantum
interference. This fact is clearly seen if the density matrix is
represented by the Wigner function. On the torus the Wigner function
exhibits two different types of interference.  The stretching and
folding produce quantum interference between different parts of
the extended state. Additionally, the periodicity of the torus
introduces interference between the state and its images. In
\fig{fig:onestepw} we show the difference between the unitary and
the noisy evolution of a coordinate state by a nonlinear map. The
two types of interference are clearly seen. The long wavelength
fringes on the convex side are produced by nonlinearities. The
short wavelength fringes correspond to the images. The effect of
the noise can be seen on (\fig{fig:onestepw}(b)) : the classical
part of the state (in white) has been broadened and the long
wavelength interference has been significantly erased. This
process continues at each step of the propagation and the quantum
state becomes more and more mixed and more and more classical.
\subsection{Spectrum of the quantum coarse gained  propagator}
    \label{sbsec:spect}
In this section we study the general features of the spectrum of
the combined action of the unitary map and the coarse graining
operator, given by (\equa{eq:superop1}). For finite values of
$\epsilon$ and $N$, $\doll$ is a convex sum of unitary matrices
and is therefore a completely positive, {\em contracting
\/} superoperator. Its spectrum has the following properties:
\begin{itemize}
 \item It is {\em unital\/} \ie it has a trivial non-degenerate
 unit eigenvalue corresponding to the uniform density
 $\hrho_\infty=\II/N$.
 \item The remaining spectrum is entirely
 contained inside the unit circle and symmetric with respect to the
 real axis. The pair of complex conjugate eigenvalues correspond to
 hermitian conjugate eigenoperators.
 \item As $\sqrt{(\doll \circ
 \doll^\dagger)}=\De$ the eigenvalues of $\De$ are also the
 singular values of $\doll$. Therefore the spectrum is contained
 exactly in the annulus
 \begin{equation}
 \text{e}^{-\left({\epsilon
   N\over\pi}\right)^2} \le |\lambda_i| \le
     \text{e}^{-\frac{1}{2}\left(\epsilon
      N\over\pi\right)^2\left(\sin^2[\pi/N]\right)}
 \end{equation}
 In the limit of large $N$ we can thus write
  \begin{equation}
     -\left(\frac{N\epsilon}{\pi}\right)^2 \le \ln |\lambda_i| \le
      -\frac{\epsilon^2}{2}.
  \end{equation}
The singular values accumulate near the origin, thus forcing most
of the eigenvalues of $\doll$ to be near zero. On the other hand
the allowed eigenvalue region extends exponentially close to the
unit circle in the limit $\epsilon \to 0$. \item The superoperator
is not normal, and therefore has distinct left and right
eigenoperators corresponding to each eigenvalue. The left and
right eigenvalue problems are then posed as follows for each pair
of complex conjugate eigenvalues $\lambda , \lambda^{*}$
\begin{eqnarray}
\doll\hR_i=\lambda_i\hR_i &
\doll\hR_i^\dagger =\lambda_i^*\hR_i^{\dag}\\
\doll^{\dagger}\hL_i=\lambda_i^{*}\hL_i &
        \doll^{\dagger}\hL_i^\dagger =\lambda_i\hL_i^\dagger
\end{eqnarray}
where $\hL_i,\, \hR_i$ conform a biorthogonal set
\begin{equation}
    \label{eq:onrelat}
\tr\left(L_i^\dagger R_j\right)=\tr\Big(L_i R_j\Big)=
      \tr\left(L_i^\dagger R_j^\dagger\right)  =   \delta_{i,j} .
\end{equation}
and we assume the normalization
 $\tr\left(L_i^\dagger L_i\right)=\tr\left(R_i^\dagger R_i\right)=1$
In particular, corresponding to $\lambda_0=1$ we choose
$\hL_0=\hR_0=\II/N$ and therefore all the remaining eigenoperators
are traceless.
  \item
 The spectral decomposition of $\doll$ then becomes
 \begin{equation}
     \doll = \sum_{i}\vket{R_i}\lambda_i\vbra{L_i}|.
 \end{equation}
\end{itemize}

The exact numerical calculation of the spectrum is hampered by the
need to diagonalize very large non-hermitian  matrices of dimension
$N^2\times N^2$ for values of $N$ large enough to extract
semiclassical features from the spectrum. In Section
{\ref{sec:method} we develop a method, specially adapted
to chaotic systems, that takes account of the dynamics of the map
to extract the part of the leading spectrum relevant to asymptotic
time behavior.

\section{Quantum classical correspondence}
    \label{sec:correspondence}
Chaotic evolution in phase space implies exponential
stretching and squeezing of initially localized densities. On a
timescale of the order of the Ehrenfest time $t_\hbar$,
significant quantum corrections to the classical evolution
inevitably appear. However, essentially classical features emerge
from quantum chaotic dynamics when decoherence is introduced, even
in the limit of no decoherence. In this section we relate
the spectra
of the propagators of densities (both classical and quantum) with
the underlying, mainly asymptotic, behavior of time dependent quantities.

Consider the classical analog for the
propagation of densities in phase space.
If $f(\nx{x})$ is a classical map, and $\nx{x}=(q,p)$ a point in
phase space, then  the
evolution of a probability density is governed by
\begin{equation}
\rho'(\nx{y})=\int\delta(\nx{y}-f(\nx{x}))\rho(\nx{x})\dif\nx{x}=
   \left[\FP\rho\right](\nx{y})
\end{equation}
where $\nx{y}=(q',p')$ and 
$\FP$ is the Perron-Frobenius (PF) operator\cite{PF}. It is unitary on the
space of square integrable functions $\Ldos$, and infinite
dimensional. However, one is mostly interested in the decay
properties of observables  much smoother than $\Ldos$.
When the functional space on which $\FP$ operates is restricted by
smoothness, the spectrum of PF changes drastically, moving to the
{\em inside\/} of the unit circle. This smoothing can
be attained by  convolution with a self adjoint compact (on
$\Ldos$) coarse graining operator $\cDe$ \cite{nonn,cgsup}, where
$\epsilon$ is the coarse graining parameter. The coarse grained
PF takes the form
\begin{equation}
    \label{eq:cgPF}
\FPe=\cDe\circ\FP,
\end{equation}
(notice the analogy with \equa{eq:superop}). $\cDe$
damps high frequency modes in $\Ldos$ and thus
effectively truncates $\FP$ to a nonunitary operator. There is
substantial difference, however, between the spectrum of the PF
for a regular map and for a chaotic  map. As the coarse graining
$\epsilon$ tends to zero, parts of the spectrum of $\FPe$ for a regular map
can be arbitrarily close to the unit circle. On
the contrary, for a chaotic map there is a finite gap for any
value of $\epsilon>0$. The isolated eigenvalues which remain
inside the unit circle as $\etoO$ are the Ruelle-Pollicot
resonances. Rugh\cite{rugh} and more recently Blank,
\etal\cite{blank} made formal descriptions of the spectrum of PF
for Anosov maps on the torus using tailor-made Banach spaces
adapted to the dynamics. Moreover Blank, \etal use this to analyze
resonances of noisy propagators and prove that these resonances
are {\it stable}, \ie independent of $\epsilon$ in the limit of
small coarse-graining . Blum and Agam\cite{agam} proposed a
numerical method to approximate the classical spectrum using
similar concepts.

A formal and very thorough recent work by Nonnenmacher\cite{nonn}
explores the characteristics of propagators, both classical and quantum, with
noise for maps on the torus, both
regular and chaotic.
In that work its is proved that,
in the limit $\NtoOO$,
the spectrum of the coarse grained quantum
propagator $\doll$, for fixed $\epsilon$, tends to that of the
coarse grained PF $\FPe$(\cite{nonn} Theorem 1). These two
theorems, taken together, provide a solid framework for the
numerical calculation of quantum resonances of torus maps and of
their classical manifestations.
\subsection{Asymptotic behavior}
    \label{sec:assymp}
The time evolution the von Neumann entropy was used by Zurek and
Paz\cite{zurek} to characterize quantum chaotic systems. They
conjectured that the rate of increase of the von Neumann entropy
of the decohering (chaotic) system is independent of the strength
of the coupling to the environment and  is ruled by the Lyapunov
exponents. Thus  classicality emerges naturally and correspondence
even for chaotic systems is recovered when decoherence is
considered. This assertion was extensively tested numerically
\cite{monte,bian,garma,patta} mainly for the linear entropy
(closely related to the purity) which is a lower bound of the von
Neumann entropy. Other quantities, like the Loschmidt
echo\cite{losch} which also display a noise independent Lyapunov decay.
have also become of interest  recently, especially in the context
of quantum information processing and computing. Besides the
linear entropy, in this section we study the asymptotic behavior
of the autocorrelation function and the Loschmidt echo.

For purely chaotic systems, after the initial spread governed by
the Lyapunov exponent, a state $\hrho$ evolved $n$ times
approaches asymptotically $\hrho_\infty=\II/N$ and all time
dependent quantities saturate to a constant value. The rate at
which these quantities saturate is given by the largest
eigenvalue, in modulus, smaller than one. Since, according to
\cite{nonn}  the spectrum of $\doll$ approaches that of $\FPe$,
then the universality of these decays can also be used to
characterize quantum chaos.

To display the decay towards $\hrho_\infty$ we subtract it from the initial state.
Thus given an arbitrary state $\hrho$, we define
\begin{equation}
    \label{eq:rhocero}
\hrho_0=\hrho-{\II\over N},
\end{equation}
were it is clear that $\tr(\hrho_0)=0$.
Thus in all computations
instead of evolving an initial {\em state\/}, we evolve an initial
traceless {\em pseudo-state\/} such as the one defined in
\equa{eq:rhocero}, orthogonal to $\hrho_\infty$. Thus, we study how the 
{\em distance\/}
between the initial state and the equilibrium state evolves. For example, for
the linear entropy, after the initial
Lyapunov behavior, which ends at about the Ehrenfest time ($n_E\sim\ln N$),
instead of saturation to the equilibrium state $\rho_\infty$, we expect to get
an unbound growth which represents how this distance decreases exponentialy, and
the exponent is proportional to $|\lambda_1|$. 

\begin{figure*}[!htb]
\begin{center}
\begin{turn}{-90}
\includegraphics*[237,194][373,599]{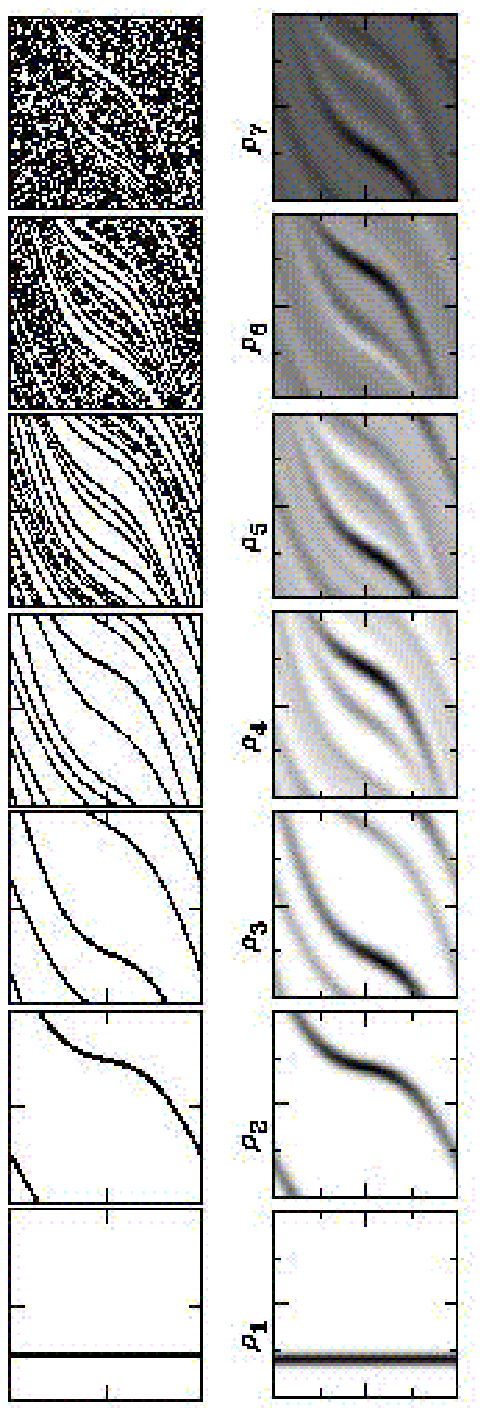}
\end{turn}
\caption{Quantum-classical correspondence for the noisy
propagator. The top row shows repeated applications of the
Perron-Frobenius operator of the perturbed Arnold cat of
\equa{eq:pcat}, to an initial classical (position) state. The
bottom row shows the Husimi representation of
$\rho_0,\ldots,\rho_6$, where $\rho_0$ is a position eigenstate
($N=150$, $\epsilon=0.2$, $k=0.02$). \label{manifolds} }
\end{center}
\end{figure*}
Assuming for simplicity that all the eigenvalues
are nondegenerate, and that $\hR_i$, $i=0,\ldots,N^2-1$,
are the right  eigenfunctions (see App.~\ref{ap:vec})
then the expansion of $\hrho_0$ in terms of $\hR_i$ is
\begin{equation}
    \label{eq:rhoexp}
\hrho_0=
\sum_{i\neq 0}r_i\hR_i,
\end{equation}
were $r_i=\tr(\hL_i^\dag\hrho_0)$ and $\hL_i$ is the left eigenfunction.
The pseudo state $\hrho_0$
evolved $n$ times, is given by
\begin{equation}
    \label{eq:decomp}
\hrho_n     =\doll^n\hrho_0=\sum_{i\neq 0}r_i\lambda_i^n\hR_i.
\end{equation}
If the eigenvalues are ordered decreasingly, according to
$1>|\lambda_1|\geq|\lambda_2|\geq\ldots\geq\lambda_{N^2-1}$, then $\hrho_0$ is a
sum of exponentially decaying modes.
Suppose  that $\lambda_1$ is real\footnote{In all the numerical
simulations made, this was indeed the case.},
then it is clear from \equa{eq:decomp} that
\begin{equation}
    \label{eq:rhotoOO}
\hrho_n \to r_1 \lambda_1^n \hR_1
\end{equation}
as $\ n\to\infty$.
Hence the asymptotic decay to the uniform
density is ruled by $\lambda_1$.
As a consequence any  quantity which depends explicitly on $\hrho_n$
shows an exponential decay.
Such is the case for the autocorrelation function
\begin{equation}
C(n)= \tr(\hrho_0^\dag\hrho_n)
\end{equation}
From \equa{eq:rhotoOO} we get,
for large $n$,
\begin{equation}
C(n)\to |r_1|^2\lambda_1^n+\ldots
\end{equation}
where  we used the fact that $\tr(\hR_1^\dag\hR_1)=1$. 
If $\lambda_1$ is complex then
\[\hrho\sim\lambda_1^nr_1\hR_1+\lambda_1^{*^n}r_1^*\hR_1,\]
and C(n) oscillates around $\lambda_1^n$ (oscillation also appears if, for example,
$|\lambda_2|\approx|\lambda_1|$).
Similarly, we can see that the linear entropy
\begin{equation}
    \label{eq:purity}
S_n=-\ln\left[\tr(\hrho^2_n)\right]
\end{equation}
grows linearly with $2n$.
Once again, using \equa{eq:rhotoOO},
the linear entropy for large $n$ is
\begin{eqnarray}
S_n&\sim&-\ln\left[|r_1|^2 |\lambda_1|^{2n} \tr\left(\hR_1^\dag\hR_1\right)\right]\nonumber \\
   &=&-2 n\ln\left[|\lambda_1|\right] + \text{\small constants}  .
\end{eqnarray}

Recently the Loschmidt echo has been extensively
studied\cite{losch} especially in the context of fidelity decay in
quantum algorithms\cite{prosen}. The definition of the echo is
\begin{equation}
    \label{eq:echodef}
M(t)=|\brac{\psi(0)}\text{e}^{{i\over\hbar}(H+\Sigma)t}
\text{e}^{-{i\over\hbar}Ht}\ket{\psi(0)}|^2
\end{equation}
which is the return probability of a state evolved forward a time
$t$ with a Hamiltonian $H$ and backward with a slightly perturbed
Hamiltonian $H+\Sigma$. It can also be viewed as the overlap
between two states  evolved forward with slightly different
Hamiltonians. Then $M$ is just a measure of how fast the two
states ``separate''. Most works focus on short times where several
``universal'' regimes have been identified. In particular noise
independent Lyapunov decay is observed for chaotic systems.

In terms of the density operator, and discrete time systems, 
the Loschmidt echo after $n$ steps is
\begin{equation}
    \label{eq:echouni}
M_n=\tr\left[\hrho_n' \hrho_n\right]=
\tr\left[\left(\SOp{U'}{^\dag}\right)^n(\hrho_0)
\SOp{U}^n(\hrho_0)\right].
\end{equation}
Where the prime represents a slight difference in the map.
If the propagation occurs in a noisy environment, characterized by $\De$, it is
natural to define the echo as
\begin{equation}
    \label{eq:echodif}
M_n(\epsilon)=\tr\left[\left(\doll^{'\dag}\right)^n(\hrho_0)\doll(\hrho_0),\right]
\end{equation}
were \equa{eq:echouni} is recovered by making $\epsilon=0$.

Following the same arguments used for the autocorrelation function and
for the linear entropy it can be shown that asymptotically
\begin{equation}
\ln\left[M_n\right]\sim n\left[\ln(|\lambda'_1|)+\ln(|\lambda_1|)\right].
\end{equation}
Notice that Schwartz inequality implies that
\begin{equation}
    \label{eq:schwartz}
\tr\left[\hrho'_n\hrho_n\right]
\leq\sqrt{\tr\left[\left(\hrho'_n\right)^2\right]
\tr\left[\left(\hrho_n\right)^2\right]}
\end{equation}
Taking the natural logarithm of the expression above we get
\begin{eqnarray}
\ln\left[M_n\right]&\leq&
\frac{1}{2}\left(\ln\left(\tr\left[\left(\hrho'_n\right)^2\right]\right)+
\ln\Big(\tr\left[\left(\hrho_n\right)^2\right]\Big) \right)\nonumber \\
\Rightarrow\ \ln\left[M_n\right]&\leq& -\frac{1}{2}\left(S'_n+S_n\right).
\end{eqnarray}
So we can see that the decay of the Loschmidt echo 
is bounded by the negative value of the
average between the linear entropy of the original system and the perturbed one
(see Fig. 4 in \cite{garma}).

These three examples illustrate the fact that in the regime where
the leading spectrum of $\doll$ and $\FPe$ coincide. We then expect
all time dependent quantities to decay asymptotically with
classical decay rates.
\subsection{Leading spectrum. Dynamics approach.}
        \label{sec:method}
In this section we describe the method used in \cite{garma} to
compute the relevant eigenvalues of the coarse grained propagator.
This method works well for hyperbolic automorphisms of $\TT$
because the nontrivial spectrum of the propagator lies entirely
inside the unit circle for all values of $\epsilon$. The existence
of a gap between 1 and $\lambda_1$ is crucial.

In any complete basis, a superoperator such as $\doll$ acting on
$\hilNN$ has associated an $N^2\times N^2$ dimensional matrix. For
small $N$ this represents no setback. However, in order to
establish a relationship between quantum and classical we need to
consider the semiclassical limit $\NtoOO$ and the diagonalization
becomes unmanageable. To overcome this problem, we use an approach
which takes advantage of the dynamics of the map to compute an
approximation of the most relevant part of the spectrum by
reducing {\em sensibly\/} the size of the eigenvalue equation.

Following \cite{rugh, blank,agam} we construct two sets ${\cal
F},\,{\cal B}\in\hilNN$ which are explicitly adapted to the
dynamics of the map \footnote{See Florido, \etal \cite{florido} 
for a rigorous review on numerical
methods that can be used to find RP resonances. The 
method used in \cite{agam}, as well as its limitations, is analized there.}.
Let $\hrho_0$ be an arbitrary initial density
in $\hilNN$, which for convenience we choose it to be a pure state
(projected onto some space). Then, by repeated application of
$\doll$ we generate
\begin{eqnarray}
    \label{eq:subspace}
{\cal F}&=&\{\hrho_0,\uhrho_1,\ldots,\uhrho_n,\ldots\},\\
{\cal B}&=&\{\hrho_0,\shrho_1,\ldots,\shrho_n,\ldots\}
\end{eqnarray}
where
\begin{eqnarray}
\uhrho_n&=&\doll(\uhrho_{n-1})=\doll^n(\hrho_0)\\
\shrho_n&=&\doll^\dagger(\shrho_{n-1})=\doll^{\dagger^n}(\hrho_0).
\end{eqnarray}
Notice that $\doll^\dagger$ is the back-step propagator.
Therefore, if the dynamics is chaotic, $\uhrho_n$ and $\shrho_n$
are increasingly smooth along the unstable and stable (classical)
directions respectively. Thus they reflect the expected behavior
of the left and right eigenfunctions of $\doll$ (see
\fig{manifolds}).

Using the bra-ket notation described in App.~\ref{ap:vec},
we  now construct the matrix
\begin{equation}
        \label{eq:Lij}
\left[\doll\right]_{i,j}=\vbra{\rho^s_i}|\doll\vket{\rho^u_j}=
\vbra{\rho^s_i}\vket{\doll(\rho^u_j)}=\vbra{\rho^s_i}\vket{\rho^u_{j+1}},
\end{equation}
where
$\vbra{\rho_i^s}|=\vbra{\doll^{\dagger^i}(\hrho_0)}|=\vbra{\rho_0}|\doll^i$.
Then we build the matrix of overlaps between elements of ${\cal
F}$ and ${\cal B}$,
\begin{equation}
    \label{eq:overlap}
\SOp{O}_{ij}=\vbra{\rho^s_i}\vket{\rho^u_j}.
\end{equation}
Notice that the structure of the matrices is very simple
\begin{equation}
\begin{array}{cclcl}
\vbra{\rho^s_i}\vket{\rho^u_j}&=&\vbra{\rho_0}|\doll^i\vket{\rho^u_j}&=&
\vbra{\rho_0}\vket{\doll^i(\rho^u_j)}\\
    &=&\vbra{\rho_0}\vket{\rho^u_{j+i}}&=&
    \vbra{\doll^{\dagger^j}(\rho^s_i)}\vket{\rho_0}\\
    &=&\vbra{\rho^s_{i+j}}\vket{\rho_0}& &
\end{array}
.
\end{equation}
We remark that $\hrho_0\in\{\hrho_\infty\}^\perp$. Because by
construction $\doll$ is trace preserving, successive applications
on an arbitrary $\rho_0$ remain in $\{\rho_\infty\}^\perp$ and
therefore the eigenvalue 1 is explicitly excluded from our
calculations. Moreover, the matrix elements in (\ref{eq:Lij}) and
(\ref{eq:overlap})decay very rapidly, providing a natural cutoff
$n_{\text{max}}$ to the sets ${\cal F}$ and ${\cal B}$.
\begin{figure}[!htb]
\scalebox{0.55}{
\includegraphics*[94,108][508,337]{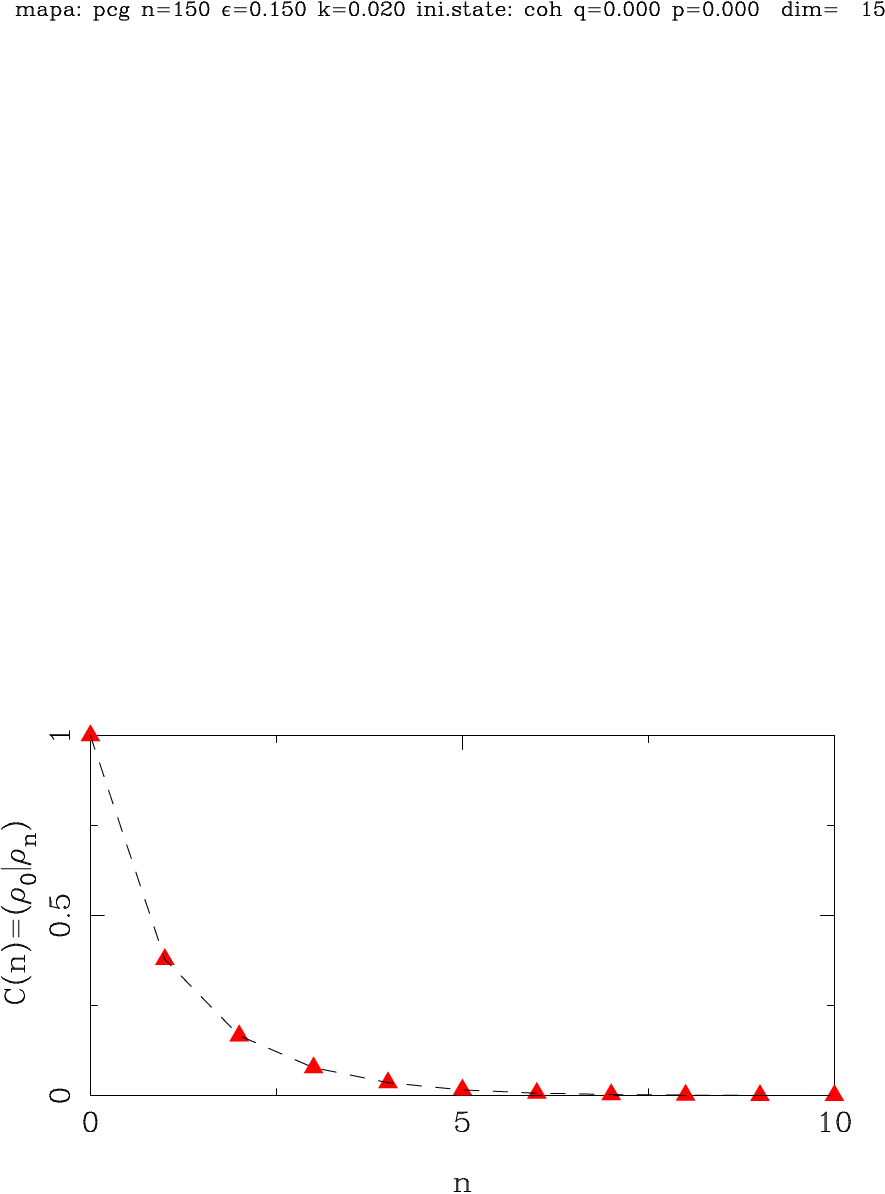}
}
\caption{ Plot of the  matrix elements
$\SOp{O}_{ij}=\text{\textbf{\textsf{L}}}_{\epsilon_{i(j-1)}}$,
where $j+i=n$. They are closely related to the
autocorrelation $C(n)=\vbra{\rho_0}\vket{\rho_n}$,
Exponential decay is observed. 
The initial state is $\hrho_0=\ket{0,0}\brac{0,0}$ where
$\ket{0,0}$ is the coherent state centered at $(0,0)$, which is a
fixed point of the map. \label{fig:corr}}
\end{figure}
In App.~\ref{app:teor} we show that an approximation of the
$\nmax$ leading eigenvalues of $\doll$  arises from the solution
of
\begin{equation}
    \label{eq:eigen}
\mbox{Det}[\left[\doll\right]_{i,j}-z\,\left[\SOp{O}\right]_{i,j}]=0,
\end{equation}
$i,j=0,\nmax-1$. This method resembles the Lanczos iteration
method \cite{golub} that uses Krylov matrices.

The combination of small matrix computations plus a strong dependence
on the dynamics makes this method a very efficient tool to get an
approximation of the leading spectrum of $\doll$ for chaotic maps.

Even when some of the main advantages of this method are evident
(reduced size, leading spectrum and spectral decomposition, etc.),
some drawbacks should be pointed out. When the classical system is
nearly integrable some resonances can remain close to the unit
circle and become unitary in the $\etoO$ limit and therefore
convergence of the method with small matrices becomes problematic.
Moreover, in that case there is a strong dependence in the initial
state $\hrho_0$. If it lies in a regular island it will not
explore all phase space. On the other hand, if initialized in the
chaotic region it will only explore the chaotic sea, leaving out
the regular tori. As a consequence some part of the relevant
spectrum is inevitably lost. Therefore, the method is useful when
the classical dynamics is fully chaotic.
\section{Numerical results}
    \label{sec:numres}
To illustrate our approach we utilize the Arnold cat map
\cite{hannay} with a small sinusoidal perturbation. The map is
\begin{equation}
  \label{eq:pcat}
  \begin{array}{rl}
    p'=&p+q\ -2 \pi k\sin [2\pi q] \\
    q'=&q+p'+2 \pi k\sin [2\pi p']
  \end{array}\ \mbox{(mod 1)}
\end{equation}
\begin{figure*}[!htb]
\begin{center}
 \scalebox{0.8}{
\includegraphics*[53,85][561,670]{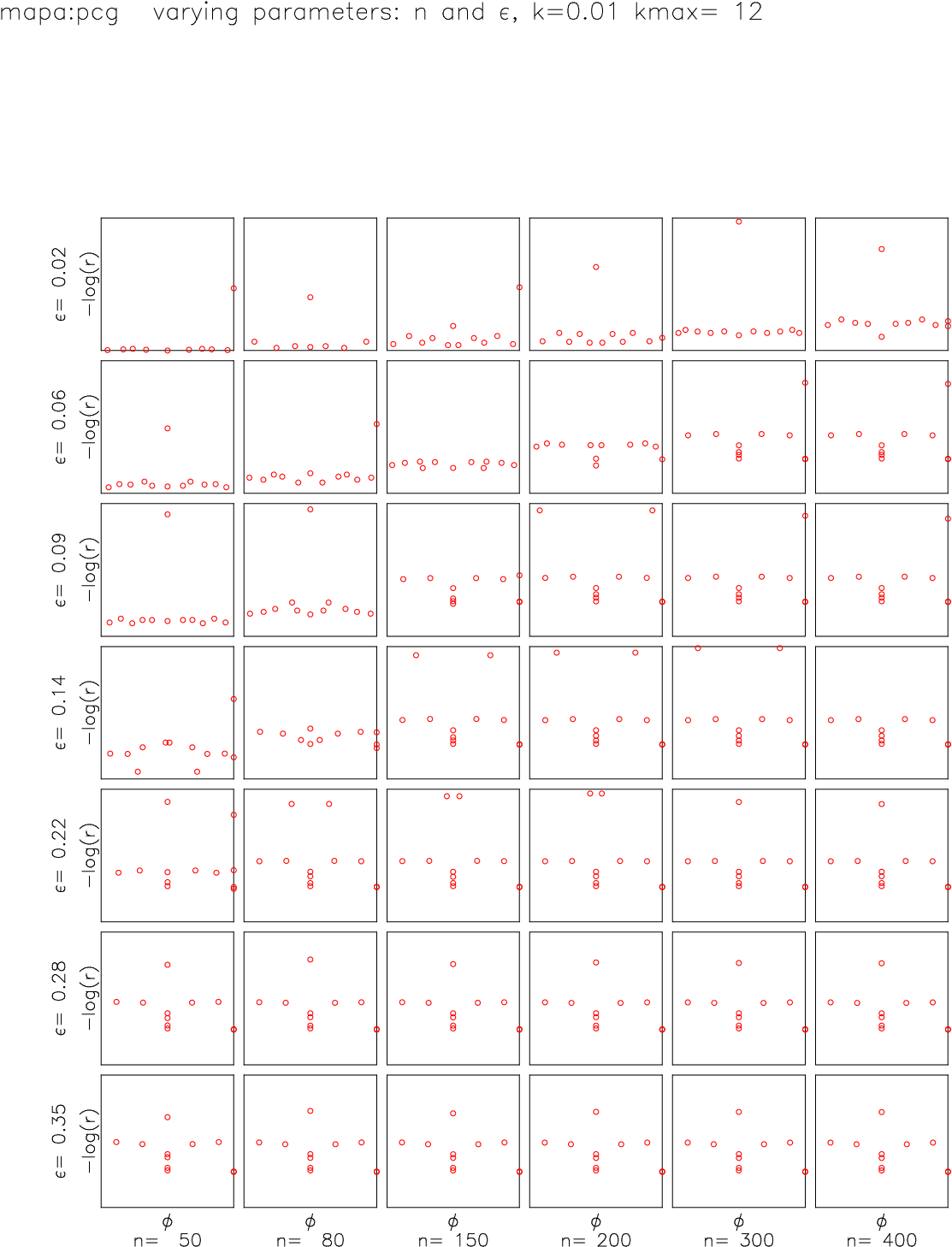}}
\caption{Leading spectrum of $\doll$ for different values of
$\epsilon$ and $N$. If $\lambda_i$ is the $i$-th eigenvalue, then
$\log\lambda_i=\log(r_i)+\rmi \phi_i$ (where $r_i=|\lambda_i|$)
and the coordinates in the plots are
$(\phi,-\log(r))$. The ranges of the axes are $\phi\in[-\pi,\pi]$ 
and $-\log(r)\in[0,6]$. 
The map is the PAC with $k=0.02$ and the
matrix was truncated to dim$=12$. \label{fig:spectra}}
\end{center}
\end{figure*}
where $k$ is the small perturbation parameter. The map has a
Lyapounov exponent which is almost independent of the value of $k$
and equal to $\lambda=\ln[(3+\sqrt{5})/2]$. On the other hand the
Ruelle resonances (computed numerically) are very sensitive to it.
Thus it is the ideal model to test the asymptotic results,
independently of the short time Lyapunov regime. The map is a
composition of two nonlinear shears and therefore it is easily
quantized as a product of two noncommuting unitary operators. The
explicit expression in the mixed representation is
\begin{eqnarray}
\brac{p}\hU\ket{q}=\exp\left\{\rmi
\frac{2\pi}{N}\left[\frac{q^2}{2}+qp-\frac{p^2}{2}\right]\right\}\times
\mbox{\hspace{.75cm}}\nonumber\\
\mbox{\hspace{0.25cm}}\times\,\exp\Big\{2\pi N\,k\Big(\cos[2\pi
q/N]+\cos[2\pi p/N]\Big)\Big\}
\end{eqnarray}
The other advantage of using a map of this type is that the
propagation both of pure states and of density matrices can be
done by fast Fourier techniques, thus allowing relatively large
Hilbert spaces with reasonable CPU times. The minor disadvantage
is that the quantization for this particular map is only valid for
even values of $N$ \cite{hannay}.
\subsection{Spectrum}
In Ref. \cite{garma} we have performed the classical calculation
of resonances and shown that the quantum and classical leading
spectra coincide. Here we take a slightly different approach and
just compute the quantum spectra for a range of $\epsilon$ and $N$
values, as shown in \fig{fig:spectra}. Observe that there is an
extended region where the spectrum is {\em stable\/} and
independent of those parameters, signifying that the eigenvalues
are properties exclusive of the map, and therefore coinciding with
the classical resonances. It is clear from this figure that the
limits $\etoO$ and $\NtoOO$ cannot be independent. In fact, at
fixed $N$ the limit $\etoO$ restores unitarity and the spectrum
returns to the unit circle.
\begin{figure}[!htb]
\scalebox{0.45}{
\includegraphics*[146,308][464,489]{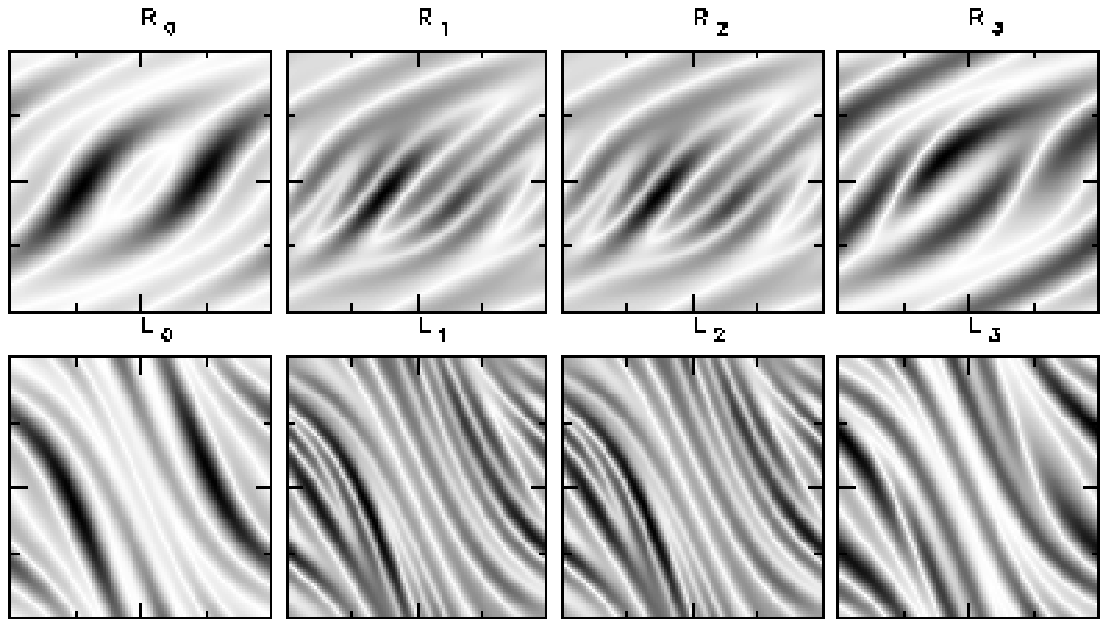}}
\caption{Top (bottom) row shows the first 4 right (left)
eigenfunctions showing the unstable (stable) manifolds for the
quantum PAC with $N=100$, $\epsilon=0.3$, $k=0.02$, and matrices
truncated to dim$=12$.\label{fig:catef}}
\end{figure}
Therefore, $\epsilon$ must decrease as a certain function of $N$.
An optimal relationship between $N$ and $\epsilon$ is yet to be
established but cannot be inferred from our limited data. However,
in our range of values a dependence like $\epsilon\sim 1/\sqrt{N}$
seems suitable.

The method described in Sect. \ref{sec:method} also provides
approximations to the eigenfunctions of $\doll$ corresponding to
the leading eigenvalues. Inside the safe region (see
\fig{fig:spectra}) of $N$ and $\epsilon$ we were able to
reconstruct at least 8 eigenfunctions successfully with matrices
of dimension of order 12. The accuracy of these eigenfunctions was
checked by evaluating the orthogonality properties in
\equa{eq:onrelat} and by computing the overlaps
\begin{equation}
\begin{array}{c}
{1\over\lambda_j}\vbrac{L_i}\doll\vket{R_j}\\
{1\over\lambda_j}\vbrac{R_i}\doll\vket{L_j}
\end{array}
.
\end{equation}
A plot of the absolute value of the Husimi representation for the
first four eigenfunctions can be seen in \fig{fig:catef}. As was
expected,  the right (left) eigenfunctions corresponding to
invariant densities of the propagator is smooth along the
classical unstable (stable) manifold of the corresponding map. The
right (left) eigenfunctions are not uniform along the unstable
(stable) manifold showing pronounced peaks at the position of
short periodic points. We intend to make a systematic analysis of
this connection in a future work.
\subsection{Asymptotic decay}
    \label{sbsec:adecay}
In this section we study numerically the asymptotic behavior of
the autocorrelation function, the linear entropy and the Loschmidt
echo for the PAC. In \fig{fig:entcorr} we see the growth of
$C_n=-\ln[C(n)]$ and the growth of $S_n$ for the perturbed Arnold
cat defined in \equa{eq:pcat} in Sect.~\ref{sec:numres}. In both
cases there are two well defined regimes. Initially both grow with
the slope determined by the Lyapunov exponent of the map. For the
PAC the Lyapunov is essentially the same for a wide range of
perturbations. On the other hand, the Ruelle resonances depend
strongly on the perturbation\footnote{see Fig. 3 in
\cite{garma}.}. 
Taking as initial density a traceless pseudo-state 
(see \equa{eq:rhocero}), time evolution of quantities show how the state 
approaches uniformity exponentially, with a rate given by the largest RP 
resonance.
We observe that, after the Lyapunov regime (around the Eherenfest time $n_E$
\footnote{In \fig{fig:entcorr} $N=450$ so
$n_E\sim\ln N=6.11$.}), the slope
of the growth of $C_n$ is given by $\ln|\lambda_1|$ whereas the slope of $S_n$ is given by
$2\ln|\lambda_1|$ as predicted.
This factor two arises from the square in the definition of $S_n$ and is
clearly seen in \fig{fig:entcorr}. The solid lines represent these two
slopes and were obtained by
computing $\lambda_1$ using the method described in Sec. \ref{sec:method}.
\begin{figure}[!htb]
\begin{center}
\scalebox{0.45}{
\includegraphics*[79,65][580,388]{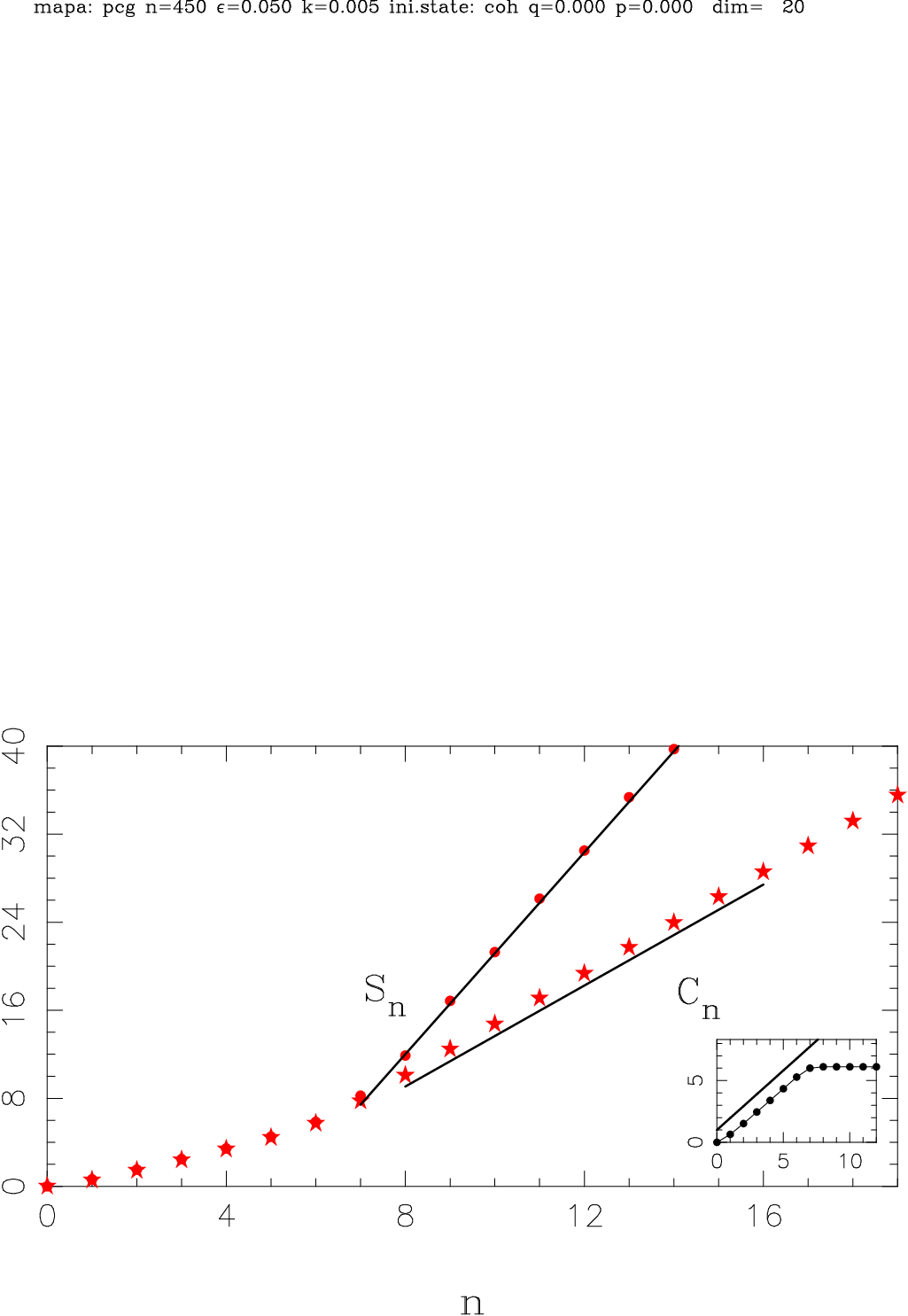}}
\caption{ Purity decay and Correlation decay for the PAC
(\equa{eq:pcat}) with $N=450$, $\epsilon=0.05$, $k=0.005$,
initial pseudo-state $\hrho=\ket{0,0}\brac{0,0}-\II/N$, where
$\ket{0,0}$ is a coherent state centered at $(0,0)$. The inset
shows the evolution of $S_n$ for $\hrho=\ket{0,0}\brac{0,0}$ and
how it saturates to the constant value $\ln
N$.\label{fig:entcorr}}
\end{center}
\end{figure}

In order to show the universality of the decay of the linear
entropy and the Loschmidt echo, in terms of classical quantities,
in \fig{fig:entroecho} we show $S_n$ and $\ln(M_n(\epsilon))$ for
various values of the parameter $\epsilon$. The linear entropy is
simply the negative logarithm of the  purity $\tr(\hrho_n^2)$.
\noindent
\begin{figure*}[!htb]
\begin{center}
\scalebox{0.4}{
\includegraphics*[46,66][585,538]{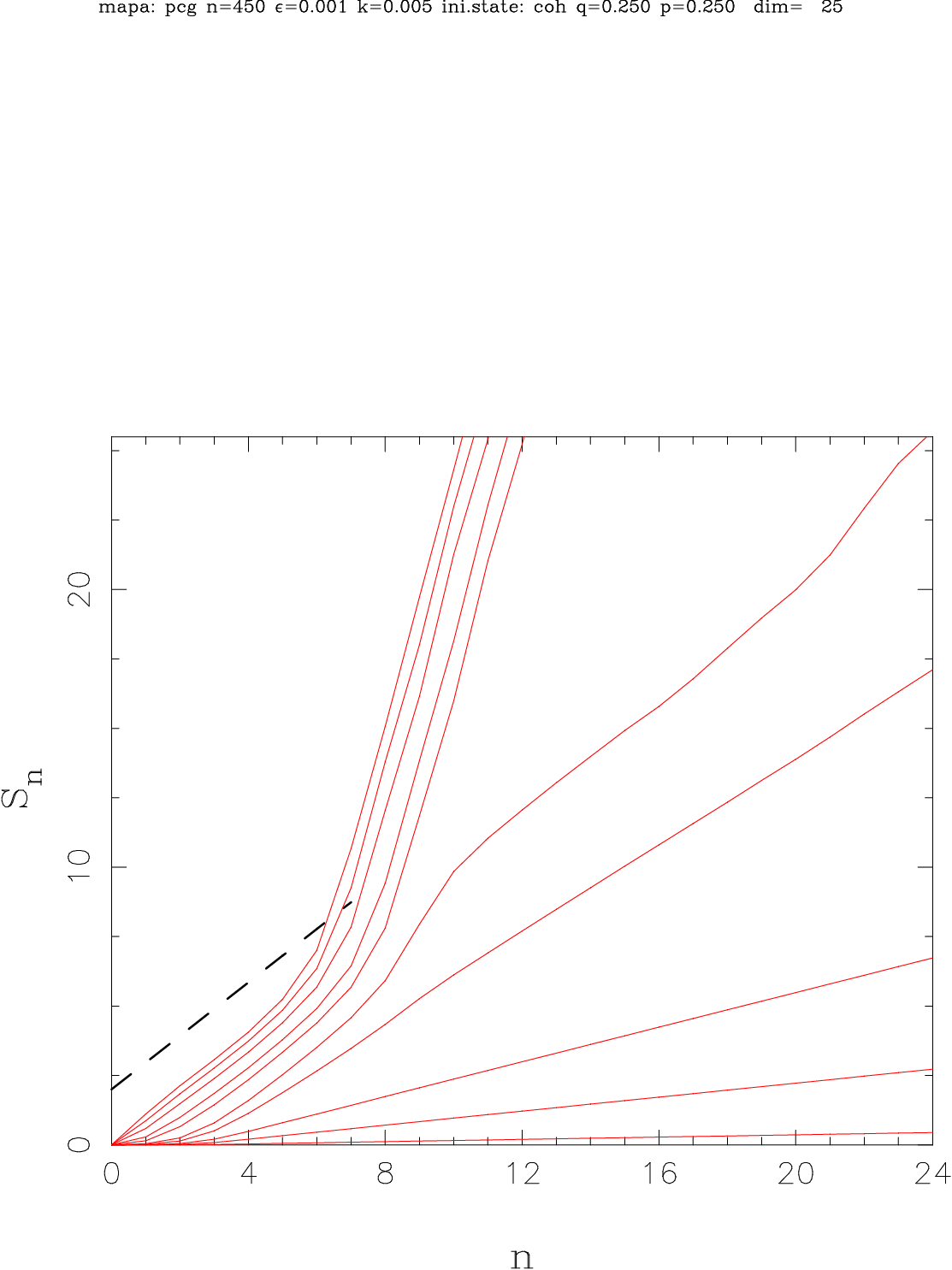}
}
\scalebox{0.4}{
\includegraphics*[44,66][575,538]{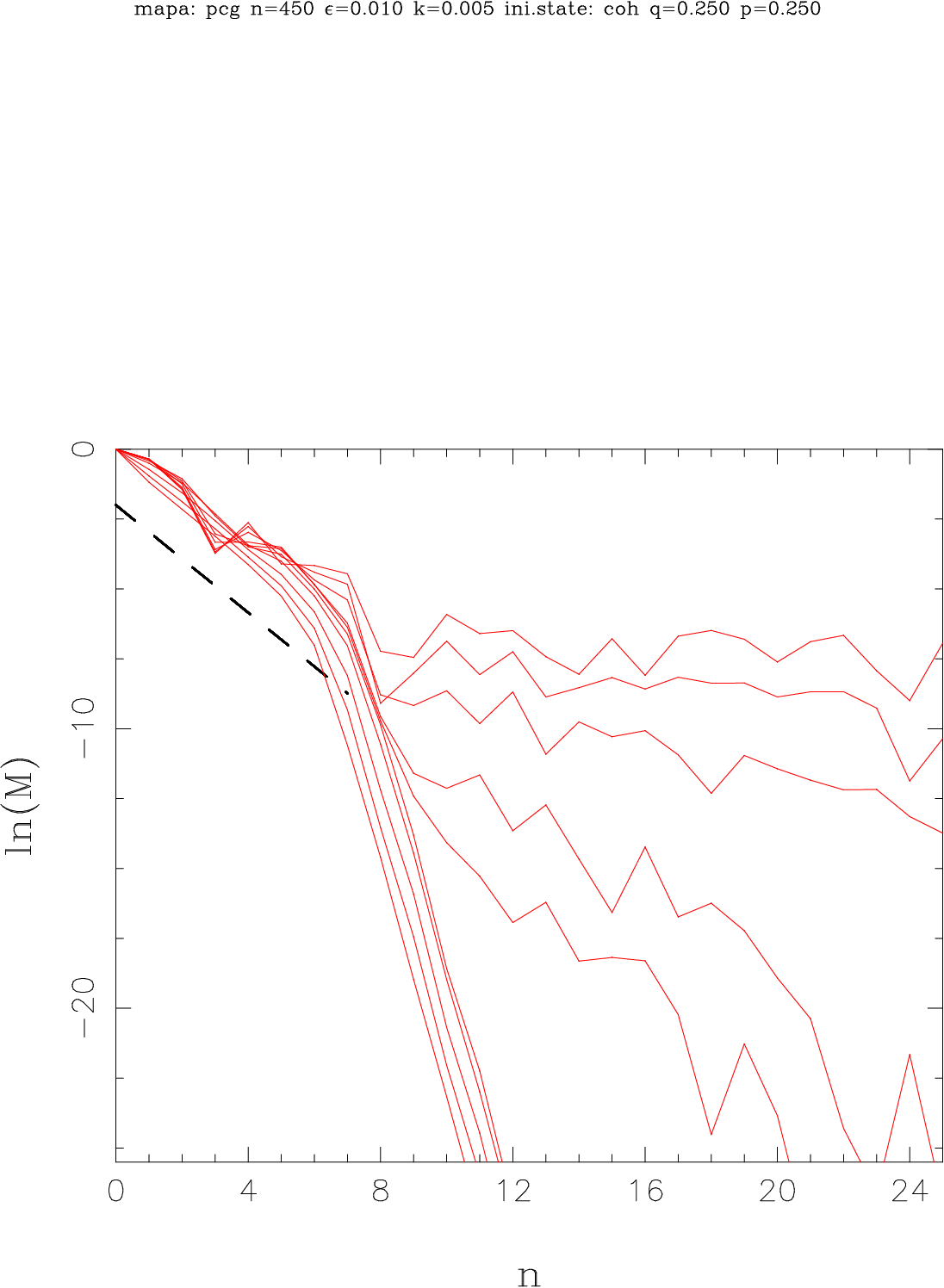}
} \caption{Linear entropy growth (left panel) and Loschmidt echo
decay (right) for various values of $\epsilon$, ranging from
$0.001$ to $0.1$ and for the PAC with $N=450$, $k=0.005$. Both
Lyapunov and Ruelle regimes can be seen when the rates saturate at
an $\epsilon$-independent value. \label{fig:entroecho}}
\end{center}
\end{figure*}
When $\epsilon\sim 0$ the purity is conserved and equal to one, so
the linear entropy does not grow. However when $\epsilon\neq 0$
the purity will decay at a rate proportional to $\epsilon$. At one
point, as predicted in \cite{zurek}, the growth of the linear
entropy saturates and no longer depends on $\epsilon$. Since we
evolved a traceless pseudo-state, with no component on the uniform
density, after the Lyapunov growth the Ruelle-Pollicot regime
appears. In the same way as for the entropy, for small values of
$\epsilon$ the asymptotic decay rate is $\epsilon$ dependent but
it saturates when rate determined by the first Ruelle resonance is
attained. This phenomenon can clearly be seen in
\fig{fig:entroecho} (left). In the right panel we display the echo
and illustrate exactly the same feature.
\section{Conclusions}
    \label{sec:conclusion}
We have developed a method to study numerically the spectral
properties of open quantum maps on the torus. The method is
particularly well suited to chaotic maps and provides reliable
eigenvalues and eigenfunctions. The noise model that we
implemented utilizes phase space translations as Kraus operators
and is equivalent to coarse graining quantum Markovian master
equations. Therefore it brings out classical properties of the map
and we have shown that these properties are reflected in the
asymptotic decay of several quantities. The same methods can be
used to study other noise models in the context of quantum
information theory, if one thinks of the quantum map as an
algorithm to be implemented and the noise as the error source
present in any implementation.

\begin{acknowledgments}
The authors profited from discussions with Eduardo Vergini, Diego Wisniacki,
Fernando Cucchietti and Stephanne Nonnenmacher. We would like to thank the
referees for useful comments as well as for pointing out Ref. \cite{florido}. 
Financial support was provided by CONICET and ANPCyT.
\end{acknowledgments}
\appendix
\section{Adjoint and linear action}
    \label{ap:vec}
Let $\hilN$ be a complex Hilbert space of dimension $N$.
The space of linear operators
acting on $\hilN$ is called Liouville space $\frak{L}\equiv\hilNN$.
Elements  in $\frak{L}$ are usually represented by
$N\times N$ dimensional complex matrices. However,
given $\hat{A},\,\hat{B}\in\hilNN$
then the ``canonical'' inner product, which induces the norm, is
\begin{equation}
(\hA,\hB)=\tr(\hA^\dagger\hB).
\end{equation}
Thus  $\hilNN$ is a Hilbert space itself. Now, superoperators are
a subset of the space of linear operators acting on $\hilNN$. We
introduce a {\em bra-ket\/} notation to simplify inner product
expressions but also to distinguish the two types of
decompositions we use for superopertors. Let $\hA,\,\hB\in\hilNN$
then the action of a superoperator $\SOp{S}:\hilNN\to\hilNN$ can
be written as
\begin{equation}
    \label{eq:ordinary}
\hB=\SOp{S}(\hA)
\end{equation}
or as
\begin{equation}
    \label{eq:leftright}
\vket{B}=\SOp{S}\vket{A}
\end{equation}
indistinctly. The adjoint, in the bra-ket form, is defined as
usual by
\begin{equation}
\vbra{A}\vket{\SOp{S}(\hB)}=
\vbra{A}|\SOp{S}\vket{B}=
\vbra{\SOp{S}^\dagger(\hA)}\vket{B},
\end{equation}
which settles that
$\vbra{A}|\SOp{S}=\vbra{\SOp{S}^{^\dagger}(\hA)}|$. Summarizing,
\begin{equation}
\begin{array}{rclcrcl}
\hA&\equiv&\vket{A}&\mbox{\hspace{0.5cm}}&
(\hA,\, \cdot\,)&\equiv&\vbrac{A}           \\
(\hA,\hB)&\equiv&\vbra{A}\vket{B}& \mbox{\hspace{0.5cm}}&
\SOp{S}(\hA)&\equiv&\SOp{S}\vket{A}
\end{array}
.
\end{equation}
One way to think about it (not absolutely necessary but helpful)
is to think of $\hA$ as an operator, or matrix, in an operator
space, acting on vectors, and $\vket{A}$ as a vector in a vector
space, acted on by superoperators.

Now, a completely positive superoperator has a Kraus operator sum
representation. Suppose $\SOp{S}$ is a completely positive
supeorperator then there exist a set of operators
$\{\hM_\mu\}_{\mu=0}^{N^2-1}\in\hilNN$, such that
\begin{equation}
\SOp{S}=\sum_{\mu}\hM_{\mu}\otimes\hM_{\mu}^{\dagger}
\end{equation}
$\hM_\mu$ are the Kraus operators. Without loss of generality, if
the number of operators is smaller than $N^2$ we can always
complete the set with zeros. The {\em adjoint\/} action of
$\SOp{S}$ on an operator $\hA$ is defined through the Kraus
representation suitable for the case of \equa{eq:ordinary}
\begin{equation}
    \label{eq:SKrauss}
\SOp{S}(\hA)=\sum_{\mu}\hM_{\mu}\hA\hM_{\mu}^{\dagger}.
\end{equation}
\equa{eq:SKrauss} determines how the Kronecker product symbol
$\otimes$ should be interpreted throughout this work.

On the other hand, a superoperator $\SOp{S}$ can be written as an
expansion of spectral projectors. Let $\hat{R}_{i}$ and
$\hat{L}_i$ be right and left eigenoperators of $\qL$
respectively, such that
\begin{equation}
\begin{array}{lll}
\SOp{S}(\hat{R}_i)&=&\lambda_i\hat{R_i}\\
\SOp{S}^{^\dagger}(\hat{L}_i)&=&\lambda_i^{*}\hat{L_i}
\end{array}
,\ \mbox{\hspace{1cm}}i=1,\ldots,N^2,
\end{equation}
and assume for simplicity that $\lambda_i$ are nondegenerate.
Then the spectral projectors are
\begin{equation}
\hat{R}_i\,\tr(\hat{L}_i^{^\dagger},\,\cdot\,)=\vket{R_i}\vbra{L_i}|,
\end{equation}
and the spectral decomposition is given by,
\begin{eqnarray}
    \label{eq:spectrald}
\SOp{S}&=&\sum_{i}\vket{R_i}\lambda_i\vbra{L_i}|\\
\SOp{S}^\dagger&=&\sum_{i}\vket{L_i}\lambda_i^*\vbra{R_i}|.
\end{eqnarray}
Therefore, given the spectral decomposition, the two ways of expressing
the action of $\SOp{S}$ on $\hA$ are
\begin{equation}
    \label{eq:bothactns}
\SOp{S}\vket{A}=\sum_{i}\vket{R_i}\lambda_i\vbra{L_i}\vket{A}\equiv
\sum_{i}\hat{M}_{\mu}\hA{\hat{M}_\mu} ^{^\dagger}=\SOp{S}(\hA)
\end{equation}
In more general terms Caves\cite{caves} identifies and describes
the two different ways a superoperator acts as {\em ordinary
action\/} (\ref{eq:ordinary}) and {\em left-right action\/}
(\ref{eq:leftright}). This provides two distinct decompositions
of the same superoperator.
\section{Leading eigenvalues of a large matrix}
    \label{app:teor}
In this section we describe in a general way the method used to compute
the leading eigenvalues of
the superoperator $\doll$ in section~\ref{sec:method}.
It is based on the Lanczos power
iteration method\cite{golub}
but was inspired by a recent work by Blum and Agam \cite{agam}.
This method is useful when only a
few of the largest (in modulus) eigenvalues is needed and also,
since it deals with large
matrices, when there is an efficient subroutine to implement
the matrix-vector product but there
is no need to store the whole matrix in an array variable. Moreover, convergence and accuracy depends
strongly on the distance part of the spectrum one wants to calculate and the part to be neglected.

In this work we don't address the question of the estimation of
errors.

\noindent
\begin{prop}
. Suppose $\grkbf{A}$ is a large, sparse matrix in $\CC^{n\times
n}$ and assume each of its eigenvalues  $\lambda_i$ has
multiplicity one and that
$1\ge|\lambda_0|>|\lambda_1|\ldots>|\lambda_{n-1}|$. Suppose
$\left\{l_i\right\}_{i=0}^{n-1}$ and
$\left\{r_i\right\}_{i=0}^{n-1}$ are the corresponding left and
right eigenvectors
\begin{eqnarray}
\grkbf{A}r_i&=&\lambda_i r_i\\
\grkbf{A}^\dag l_j &=& \lambda^*_j l_j,
\end{eqnarray}
and let
$u_0\in\CC^n$ be a vector such that
\begin{equation}
|(l_i,u_0)|>0 \ \ \mbox{and}\ \ |(r_i,u_0)|>0\ \ \forall i<k
\end{equation}
for some $k\leq n$, where $(\ ,\ )$ represents as usual the inner product.
Then the
first $k$ eigenvalues can be estimated
from the {\em reduced} ($k\times k$) eigenvalue equation
\begin{eqnarray}
\mbox{Det}\Big[\grkbf{\cal K}^T(\grkbf{A}^\dagger,u_0,k)\grkbf{A}
    \grkbf{\cal K}(\grkbf{A},u_0,k)
\mbox{\hspace{2.5cm}}\nonumber \\
    \label{eq:kryldet}
-z\grkbf{\cal K}^T(\grkbf{A}^\dagger,u_0,k)
    \grkbf{\cal K}(\grkbf{A},u_0,k)\Big]=0
\end{eqnarray}
where $\grkbf{\cal K}(\grkbf{A},u_0,k)$ is the Krylov matrix whose columns are the iterates of $u_0$,
\begin{equation}
\grkbf{\cal K}(\grkbf{A},u_0,k)=\left[u_0,\grkbf{A}u_0,\grkbf{A}^2u_0,
    \ldots,\grkbf{A}^{k-1}u_0\right]
\end{equation}
and $^T$ as usual denotes  matrix transposition.
\end{prop}

\noindent \textbf{Proof.} We sketch a rather straightforward proof
(though perhaps not entirely rigorous). The sets
$\left\{l_i\right\}_{i=0}^{n-1}$ and
$\left\{r_i\right\}_{i=0}^{n-1}$ of left and right eigenvectors of
\grkbf{A} are complete and, they can be normalized according to
\begin{equation}
(l_i,r_j)=\delta_{ij}.
\end{equation}
Therefore there exist two distinct expansions of $u_0$
\begin{eqnarray}
u_0&=&\sum_{i=0}^{n-1} \alpha_i\,r_i\\
u_0&=&\sum_{i=0}^{n-1} \beta_i \,l_i.
\end{eqnarray}
\begin{widetext}
In terms of these expansions we obtain
\begin{equation}
\grkbf{\cal K}^T(\grkbf{A}^\dag,u_0,k)\grkbf{\cal K}(\grkbf{A},u_0,k)=
\left[
\begin{array}{c}
\sum_i\beta_i\,l_i\\
\sum_i\beta_i \lambda_i^*\,l_i\\
\vdots\\
\sum_i\beta_i\lambda_i^{*^{k-1}}\,l_i
\end{array}
\right]
\
\left[\sum_j\alpha_j\,r_j,\sum_j\alpha_j \lambda_j\,r_j,
    \cdots,\sum_j\alpha_j \lambda_j^{k-1}\,r_j\right]
\end{equation}
\end{widetext}
which yields
\begin{eqnarray}
\left[\grkbf{\cal K}^T(\grkbf{A}^\dagger,u_0,k)
\grkbf{\cal K}(\grkbf{A},u_0,k)\right]_{\mu\nu}&=&
\sum_{i,j} \alpha_j\beta_i \lambda_{j}^{\nu}\lambda_{i}^{*^\mu}(l_i,r_j)
    \nonumber \\
 &=&\sum_{i} \alpha_i \beta_i\lambda_{i}^{\nu}\lambda_{i}^{*^\mu},
\end{eqnarray}
and similarly
\begin{eqnarray}
\left[\grkbf{\cal K}^T(\grkbf{A}^\dagger,u_0,k)\grkbf{A}
\grkbf{\cal K}(\grkbf{A},u_0,k)\right]_{\mu\nu}&=&
\sum_{i,j} \beta_i\alpha_j \lambda_{j}^{\nu+1}\lambda_{i}^{*^\mu}(l_i,r_j)
    \nonumber \\
&=&\sum_{i} \beta_i\alpha_i \lambda_{i}^{\nu+1}\lambda_{i}^{*^\mu}.
\end{eqnarray}
Thus, \equa{eq:kryldet} can be re-written as
\begin{equation}
    \label{eq:secular3}
\mbox{Det}\left[\sum_{i=0}^{n-1} \alpha_i\beta_i\lambda_{i}^{\nu}\lambda_{i}^{*^\mu}(\lambda_i-z)
\right]=0.
\end{equation}
If all the conditions of the proposition are met, this equation is equivalent to
the original full eigenvalue equation.
Now, since $\lambda_i$ are ordered by dereasing modulus and assuming that the
eigenvalues accumulate around zero, leaving only a few, say $k$ of them, with significant
modulus (as is the case for the maps studied in section~\ref{sec:method}) then
we can neglect
the contribution of the last $n-k$  terms in the sum\footnote{Although they can be computed, in
this work we don't provide estimations of the errors due to this truncation.}.
Thus \equa{eq:secular3} is just the determinant of the product of three $k\times k$
square matrices
\begin{equation}
    \label{eq:secular2}
\mbox{Det}[\grkbf{\Lambda}^\dagger\, \grkbf{\Xi\, \Lambda}]=
\mbox{Det}[\grkbf{\Lambda}^\dag]\,\mbox{Det}[\grkbf{\Xi}]\,
\mbox{Det}[\grkbf{\Lambda}]=0
\end{equation}
\begin{widetext}
where
\begin{equation}
\grkbf{\Lambda}=\left(
\begin{array}{ccccc}
1&\lambda_0&\lambda_0^2&\cdots&\lambda_0^{k-1}\\
1&\lambda_1&\lambda_1^2&\cdots&\lambda_1^{k-1}\\
\vdots   &\vdots   &\vdots  &\ddots& \vdots \\
1&\lambda_{k-1}&\lambda_{k-1}^2&\cdots&\lambda_{k-1}^{k-1}
\end{array}
\right);
\mbox{\hspace{0.75cm}}
\grkbf{\Xi}= \left(
\begin{array}{cccc}
\alpha_{0}\beta_{0}(\lambda_0-z)&0&\cdots&0\\
0&\alpha_{1}\beta_{1}(\lambda_1-z)&\cdots&0\\
\vdots&\vdots&\ddots&\vdots\\
0&\cdots&0&\alpha_{k-1}\beta_{k-1}(\lambda_{k-1}-z)
\end{array}
\right).
\end{equation}
\end{widetext}
The matrix $\grkbf{\Lambda}$ is a Vandermonde matrix. The determinant of a
Vandermonde matrix $\grkbf{\Lambda}_k(\lambda_0,\ldots,\lambda_{k-1})$
is given by
\begin{equation}
    \label{eq:vander2}
\mbox{Det}\left[\grkbf{\Lambda}_k(\lambda_0,\ldots,\lambda_{k-1})\right]=
\prod_{0\le i\le j\le k-1}(\lambda_j-\lambda_i).
\end{equation}
From equation \equa{eq:vander2} it can be readily seen that if the spectrum of $\grkbf{A}$
is non-degenerate then $\grkbf{\Lambda}$ is invertible. Moreover, the structure of
$\grkbf{\Lambda}$  determines $k$ because in the limit of $k$ ``too large'', $\grkbf{\Lambda}$ is
singular, at least to within computing precision.
So, using properties of the determinant in the secular equation \equa{eq:secular2}, we get
\begin{equation}
    \label{eq:sol}
\mbox{Det}[\grkbf{\Xi}]=\prod_\mu \alpha_{\mu}\beta_{\mu}(\lambda_\mu-z)=0.
\end{equation}
Since, from the hypothesis, $\alpha_{\mu}\beta_{\mu}\neq 0$ then \equa{eq:sol}
yields the desired solution, \ie the first $k$ eigenvalues of $\grkbf{A}$. \hfill $\square$

In practice, the usefulness of the method depends upon the gap $(1-|\lambda_1|)$, because it
determines how fast the terms of the sum in \equa{eq:secular2} decay.

In section~\ref{sec:method} the span of the sets ${\cal F}$ and ${\cal B}$ are just the  Krylov
spaces\cite{golub} of $\doll$ and $\doll^\dag$, and  using the present notation
\equa{eq:eigen} is
\begin{eqnarray}
\mbox{Det}\Big[
\grkbf{\cal K}^\dag(\doll^\dagger,\hrho_0,\nmax)
\doll\grkbf{\cal K}(\doll,\hrho_0,\nmax)\mbox{\hspace{2.cm}}
    \nonumber \\
- z\,\grkbf{\cal K}^\dag(\doll^\dagger,\hrho_0,\nmax)
    \grkbf{\cal K}(\doll,\hrho_0,\nmax)
\Big]=0\mbox{\hspace{0.5cm}}
\end{eqnarray}
In analogy with \equa{eq:secular2}.
The efficiency of this method depends strongly on the spectrum
configuration. The case of the coarse-grained propagator of
hyperbolic maps on the torus\cite{blank,nonn} is particularly
favorable because of the significant gap between 1 and $\lambda_1$
and because 0 is an accumulation point, so a large number of
resonances can be discarded and the size of the matrices is
reduced dramatically.

%

\end{document}